\documentclass[sigconf]{acmart}

\copyrightyear{2017} 
\acmYear{2017} 
\setcopyright{acmlicensed}
\acmConference{CIKM'17}{}{November 6--10, 2017, Singapore.}
  \acmPrice{15.00}
  \acmDOI{https://doi.org/10.1145/3132847.3133021}
  \acmISBN{ISBN 978-1-4503-4918-5/17/11}

\usepackage[utf8]{inputenc}
\usepackage{subfigure}
\usepackage{multirow}
\usepackage{color, verbatim,caption}
\usepackage{hyperref}
\usepackage{enumitem}
\usepackage{algorithm}
\usepackage{algorithmicx}
\usepackage{algpseudocode}
\usepackage{amsmath}

\usepackage{array}
\newcolumntype{L}[1]{>{\raggedright\let\newline\\\arraybackslash\hspace{0pt}}m{#1}}
\newcolumntype{C}[1]{>{\centering\let\newline\\\arraybackslash\hspace{0pt}}m{#1}}
\newcolumntype{R}[1]{>{\raggedleft\let\newline\\\arraybackslash\hspace{0pt}}m{#1}}

\begin{document}


\title{An Attention-based Collaboration Framework for Multi-View Network Representation Learning}

\author{Meng Qu$^1$, Jian Tang$^{2,3}$, Jingbo Shang$^1$, Xiang Ren$^1$, Ming Zhang$^4$, Jiawei Han$^1$}

\affiliation{%
  \institution{{$^1$}{University of Illinois at Urbana-Champaign, IL, USA}}
}
\affiliation{%
  \institution{{$^2$}{HEC Montreal, Canada}}
}
\affiliation{%
  \institution{{$^3$}{Montreal Institute of Learning Algorithms, Canada}}
}
\affiliation{%
  \institution{{$^4$}{Peking University, Beijing, China}}
}
\affiliation{%
  \institution{{$^1$}{\{mengqu2, shang7, xren7, hanj\}@illinois.edu} $\quad$ {$^{2,3}$}{tangjianpku@gmail.com} $\quad$ {$^4$}{mzhang\_cs@pku.edu.cn}}
}

\begin{abstract} 

Learning distributed node representations in networks has been attracting increasing attention recently due to its effectiveness in a variety of applications. Existing approaches usually study networks with a single type of proximity between nodes, which defines a single view of a network. However, in reality there usually exists multiple types of proximities between nodes, yielding networks with multiple views. This paper studies learning node representations for networks with multiple views, which aims to infer robust node representations across different views. We propose a multi-view representation learning approach, which promotes the collaboration of different views and lets them vote for the robust representations. During the voting process, an attention mechanism is introduced, which enables each node to focus on the most informative views. Experimental results on real-world networks show that the proposed approach outperforms existing state-of-the-art approaches for network representation learning with a single view and other competitive approaches with multiple views.

\end{abstract}


\maketitle

\section{Introduction}
\label{sec::intro}

Mining and analyzing large-scale information networks (e.g., social networks~\cite{wasserman1994social}, citation networks~\cite{sun2009ranking} and airline networks~\cite{jaillet1996airline}) has attracted a lot of attention recently due to their wide applications in the real world. To effectively and efficiently mine such networks, a prerequisite is to find meaningful representations of networks. Traditionally, networks are represented as their adjacency matrices, which are both high-dimensional and sparse. Recently, there is a growing interest in representing networks into low-dimensional spaces (a.k.a, network embedding)~\cite{perozzi2014deepwalk,tang2015line, grovernode2vec}, where each node is represented with a low-dimensional vector. Such vector representations are able to preserve the proximities between nodes, which can be treated as features and benefit a variety of downstream applications, such as node classification~\cite{perozzi2014deepwalk,tang2015line}, link prediction~\cite{grovernode2vec} and node visualization~\cite{tang2015line}.

\begin{figure}
	\centering
	\vspace{0.3cm}
	\includegraphics[width=0.48\textwidth]{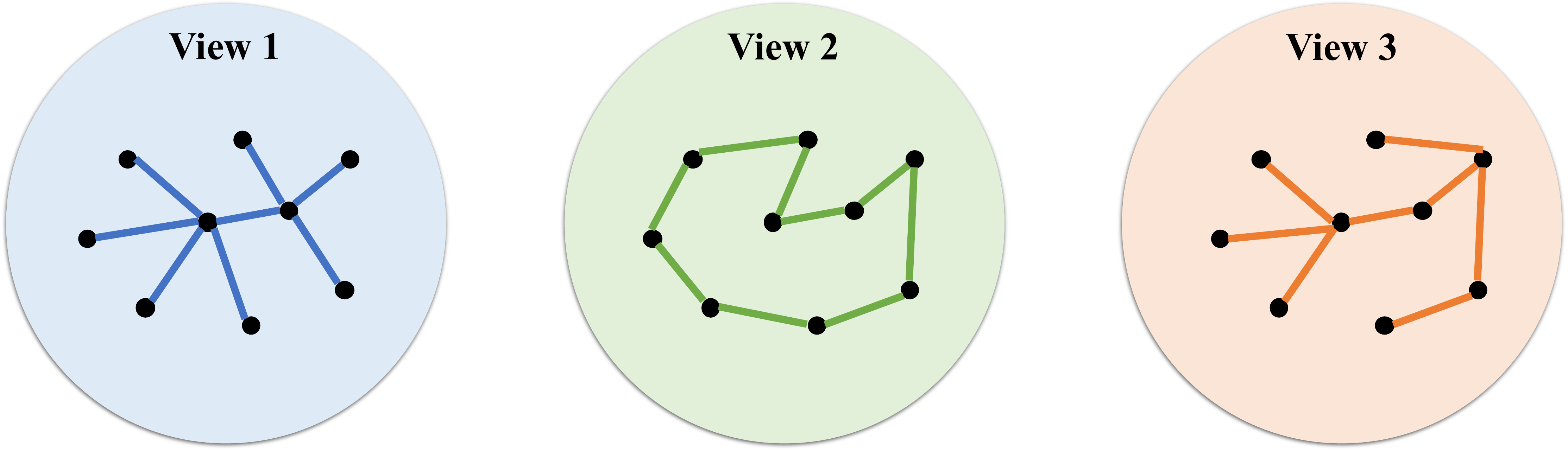}
	\caption{An example multi-view network with three views. Each view corresponds to a type of proximity between nodes, which is characterized by a set of edges. Different views are complementary to each other.}
	\label{fig::example}
	\vspace{-0.3cm}
\end{figure}

Though empirically effective and efficient in many networks, all these work assumes there only exists a single type of proximity between nodes in a network, whereas in reality multiple types of proximities exist. Take the network between authors in the scientific literature as an example, the proximity can be induced by co-authorship, meaning whether two authors have once coauthored a paper, or citing relationship, meaning whether one author cited the papers written by the other one. Another example is the network between users in social media sites (e.g, Twitter), where multiple types of proximities also exist such as the ones induced by the following-followee, reply, retweet, and mention relationships. Each proximity defines a view of a network, and multiple proximities yield a network with multiple views. Each individual view is usually sparse and biased, and thus the node representations learned by existing approaches may not be so comprehensive. To learn more robust node representations, a natural solution could be leveraging the information from multiple views. 

This motivated us to study a new problem: \emph{learning node representations for networks with multiple views}, aiming to learn robust node representations by considering multiple views of a network. In literature, various methods have been proposed for learning data representations from multiple views, such as multi-view clustering methods~\cite{kumar2011co,xia2010multiview,zhou2007spectral,chaudhuri2009multi,kumar2011co} and multi-view matrix factorization methods~\cite{liu2013multi,greene2009matrix,singh2008relational}. These methods perform well on many applications such as clustering~\cite{kumar2011co,liu2013multi} and recommendation~\cite{singh2008relational}. However, when applied to our problem, they have the following limitations: 
(1) \emph{Insufficient collaboration of views.} As each individual view of a network is usually biased, learning robust node representations requires the collaboration of multiple views. However, most of existing approaches for multi-view learning aim to find compatible representations across different views rather than promote the collaboration of different views for finding robust node representations.
(2) \emph{Lack of weight learning.} To learn robust node representations, the information from multiple views needs to be integrated. During integration, as the importance of different views can be quite different, their weights need to be carefully decided. Existing approaches usually assign equal weights to all views. In other words, different views are equally treated, which is not reasonable for most multi-view networks.
To overcome the limitations, we are seeking an approach that is able to promote the collaboration of different views, and also automatically infer the weights of views during integration.

In this paper, we propose such an approach. We first introduce a set of \emph{view-specific} node representations to preserve the proximities of nodes in different views. The view-specific node representations are then combined for voting the \emph{robust} node representations. Since the quality of the information in different views may be different, it would be ideal to treat the views differently during the voting process. In other words, each view should be weighted differently. Inspired by the recent progress of the attention mechanism~\cite{bahdanau2014neural} for neural machine translation, in this paper we propose an attention based method to infer the weights of views for different nodes, which will leverage a few labeled data. The whole model can be efficiently trained through the backpropagation algorithm~\cite{rumelhart1988learning}, alternating between optimizing the view-specific node representations and voting for the robust node representations by learning the weights of different views. 

We conduct extensive experiments on various real world multi-view networks. Experimental results on both the multi-label node classification task and link prediction task show that our proposed approach outperforms state-of-the-art approaches for learning node representation with individual views and other competitive approaches with multiple views. 

In summary, in this paper we make the following contributions:
\begin{itemize}[leftmargin=*,noitemsep,nolistsep]
	\item We propose to study \emph{multi-view network representation learning}, which aims to learn node representations by leveraging information from multiple views.
	\item We propose a novel collaboration framework, which promotes the collaboration of different views to vote for robust node representations. An attention mechanism is introduced for learning the weights of different views during voting.	
	\item We conduct experiments on several multi-view networks. Experimental results on two tasks prove the effectiveness and efficiency of our proposed approach over many competitive baselines. 
\end{itemize}

\section{Problem Definition}
\label{sec::definition}
In this section, we introduce some background knowledge and formally define the problem of multi-view network embedding. We first define information networks and their views as follows:

\begin{definition}
\label{def::View}
\textbf{(Information Network, View)}
\textsl{ An \textbf{Information Network}, denoted as $G=(\mathcal{V},E)$, encodes the relationships between different objects, where $\mathcal{V}$ is a set of objects and $E$ is a set of edges between the objects. Each edge $e=(u,v)$ is associated with a weight $w_{uv} > 0$, indicating the strength of the relationship between $u$ and $v$. A \textbf{view} of a network is derived from a single type of proximity or relationship between the nodes, which can be characterized by a set of edges $E$.}
\end{definition}

Traditionally, networks are represented as their adjacency matrices, which are sparse and high-dimensional. Recently, learning low-dimensional vector representations of networks (a.k.a.\ network embedding) attracts increasing attention, which is defined below:

\begin{definition}
	\label{def::Embedding}
	\textbf{(Network Embedding)}
	\textsl{Given an information network $G=(V, E)$, the problem of \textbf{network embedding} aims to learn a low-dimensional vector representation $\mathbf{x}_v \in R^d$ for each node $v$ with $d\ll|V|$, which preserves the proximities between the nodes. }
\end{definition}

Various network embedding approaches~\cite{tang2015line,perozzi2014deepwalk,grovernode2vec} have been proposed recently. Although they have been proved to be effective and efficient in various scenarios, they all focus on networks with a single type of proximity/relationship. However, in reality we often observe multiple types of proximities/relationships between nodes. For example, for users in social media sites such as Twitter, besides the following relationships, other relationships also exist such as retweet, meaning one user forwarded the tweets written by another user, and mention, meaning one user mentioned another user in his tweets. 
Each type of proximity/relationship defines a view of networks between nodes, and multiple types of proximity/relationship yield networks with multiple views. Different views of a network are usually complementary to each other, and thus considering multiple views may help learn more robust node representations. This motivated us to study the problem of learning node representations for networks with multiple views, and we formally define the problem as follows:

\begin{definition}
	\label{def::LMINE}
	\textbf{(Multi-view Network Embedding)}
	\textsl{Given an information network with $K$ views, denoted as $G=(\mathcal{V},E_1, E_2,\ldots, E_K)$, the problem of \textbf{Multi-view Network Embedding} aims to learn the \textbf{robust} node representations $\{\mathbf{x}_v\}_{v \in \mathcal{V}} \subseteq R^d$, which are consistent across different views. $R^d$ is a low-dimensional space with $d\ll |\mathcal{V}|$.}
\end{definition}

To learn robust node representations across different views, it would be desirable to design an approach to promote the collaboration of different views and vote for the robust node representations. Since the quality of views are different, the approach should also be able to weight the views differently during voting. In the next section, we introduce such an approach.

\section{Multi-view Network Embedding}

In this section, we introduce our proposed approach for embedding networks with multiple views. 
When applied to the problem, most existing approaches, e.g., multi-view clustering and multi-view matrix factorization algorithms, fail to achieve satisfactory results. This is because they cannot effectively promote the collaboration of different views during training. Moreover, they also cannot assign proper weights to different views when combining the information from them.

To solve these challenges, our approach first mines the node proximities encoded in single views, during which, a collaboration framework (Sec.~\ref{subsec::embedding}) is proposed to promote the collaboration of views. After that, we further integrate different views to vote for more robust node representations. During voting, we automatically learn the voting weights of views through an attention based approach (Sec.~\ref{subsec::weight}).

The overall objective of our approach is summarized below:
\begin{equation}
\label{eqn::collab}
	O = O_{collab} + O_{attn}.
\end{equation}
$O_{collab}$ is the objective function of the collaboration framework, in which we aim to learn the node proximities in individual views and meanwhile vote for the robust node representations. $O_{attn}$ is the objective function for weight learning. Next, we introduce the details of each part.

\begin{figure}
	\centering
	\includegraphics[width=0.48\textwidth]{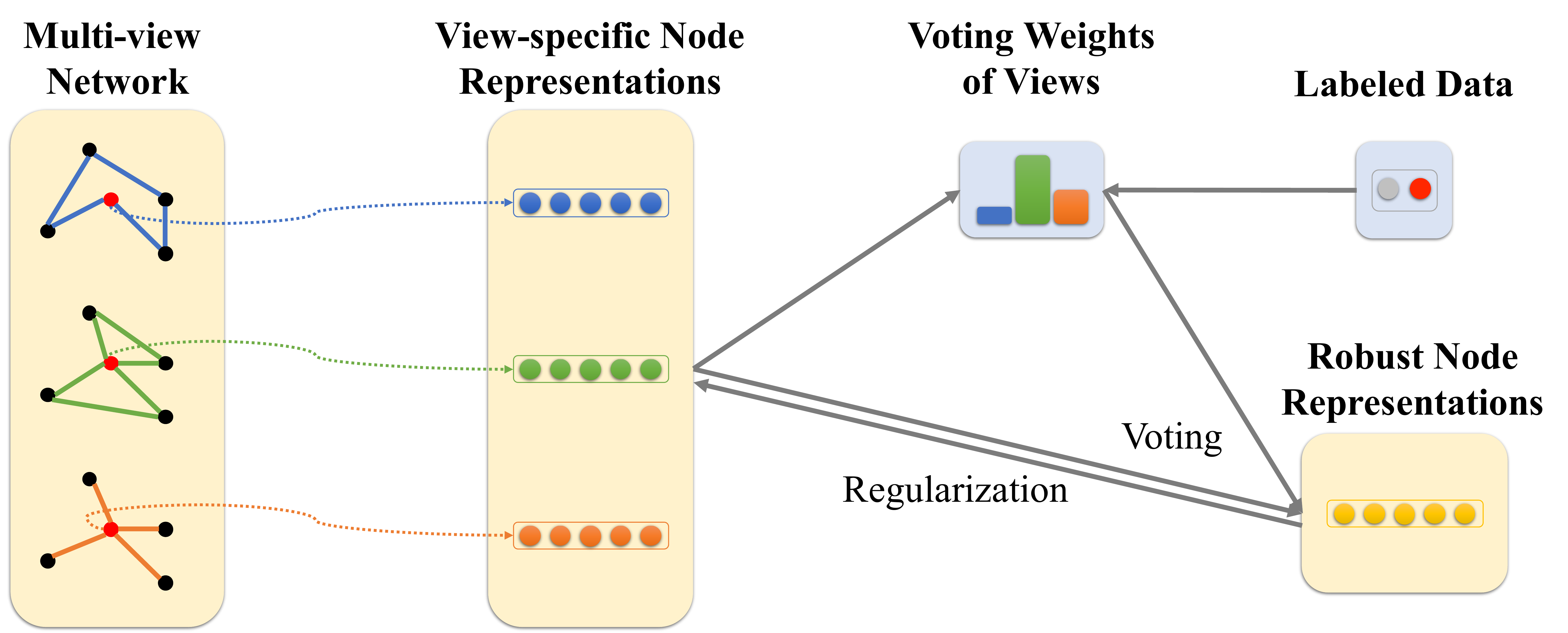}
	\caption{Overview of the proposed approach. The collaboration framework (yellow parts) preserves the node proximities of different views with a set of view-specific node representations, which further vote for the robust representations. During voting, we learn the weights of views through an attention based method (blue parts), which enables nodes to focus on the most informative views.}
	\label{fig::framework}
\end{figure}

\subsection{Collaboration Framework}
\label{subsec::embedding}
The goal of the collaboration framework is to capture the node proximities encoded in individual views and meanwhile integrate them to vote for the robust node representations.
Therefore, for each node $v_i$, we introduce a set of view-specific representations $\{\mathbf{x}_i^k\}_{k=1}^K$ to preserve the structure information encoded in individual views. We also introduce a robust representation $\mathbf{x}_i$, which integrates the information from all different views.

To preserve the structure information of individual views with the view-specific node representations, for a directed edge $(v_i, v_j)$ in view $k$, we first define the probability $p_k(v_j|v_i)$ as follows:
\begin{equation}
\label{eqn::line_prob}
p_k(v_j|v_i) \propto  \exp(\mathbf{x}_j^k \cdot \mathbf{c}_i),
\end{equation}
where $\mathbf{c}_i$ is a context representations of node $v_i$. In our approach, the context representations are shared across different views, so that different view-specific node representations will locate in the same semantic space. We also tried using different context representations for different views, and we will compare with this variant in the experiments.

Following existing studies~\cite{tang2015line,tang2015pte}, for each view $k$, we try to minimize the KL-divergence between the estimated neighbor distribution $p_k(\cdot |v_i)$ and the empirical neighbor distribution $\hat{p}_k (\cdot|v_i)$. The empirical distribution is defined as $\hat{p}_k (v_j|v_i)=w_{ij}^{(k)}/d_i^{(k)}$, where $w_{ij}^{(k)}$ is the weight of the edge $(v_i,v_j)$ in view $k$ and $d_i^{(k)}$ is the out-degree of node $v_i$ in view $k$. 
After some simplification, we obtain the following objective function for each view $k$:
\begin{equation}
\label{eqn::line-view}
	O_{k} = - \sum_{ (i,j) \in E_k } w_{ij}^{(k)} \log p_k(v_j|v_i).
\end{equation}

Directly optimizing the above objective is computationally expensive because it involves traversing all nodes when computing the conditional probability. Therefore we adopt the negative sampling techniques~\cite{mikolov2013distributed, mnih2012fast}, which modify the conditional probability $p_k(v_j|v_i)$ in Eqn.~\ref{eqn::line-view} as follows:
\begin{equation}
\label{eqn::log_ns}
\log \sigma(\mathbf{c}_j \cdot \mathbf{x}_i^k)+\sum_{n=1}^NE_{v_n\sim P_{neg}^k(v)}[\log \sigma(-\mathbf{c}_n \cdot \mathbf{x}_i^k)],
\end{equation}
where $\sigma(x)=1/(1+\exp(-x))$ is the sigmoid function. The first term maximizes the probability of some observed edges, and the second term minimizes the probability of $N$ noisy node pairs, with $v_n$ sampled from a noise distribution $P_{neg}^k(v) \propto d_v^{(k)3/4}$ and $d_v^{(k)}$ is the degree of node $v$ in view $k$.

By minimizing the objective~\eqref{eqn::line-view}, the view-specific representations $\{\mathbf{x}_i^k\}_{k=1}^K$ are able to preserve the structure information encoded in different views.
Next, we promote the collaboration of different views for voting the robust node representations. In this process, as the importance of views can be quite different, we try to assign different weights to them.
With all these in mind, we introduce the following regularization term.
\begin{equation}
	\label{eqn::MVEmbed_reg}
	R=\sum_{i=1}^{|V|}\sum_{k=1}^K \lambda_i^k ||\mathbf{x}_i^k-\mathbf{x}_i||_2^2,
\end{equation}
where $||\cdot||_2$ is the Euclidean norm of a vector, $\lambda_i^k$ is the weight of view $k$ assigned by node $v_i$. Intuitively, by learning proper weights $\{ \lambda_i^k \}_{k=1}^K$ for each node $v_i$, our approach can let each node focus on the most informative views. We will introduce how we automatically learn such weights in the next section. 
By minimizing this objective function, 
different view-specific representations will vote for the robust representations based on the following equation:
\begin{equation}
\small
	\label{eqn::MVEmbed_mstep}
	\mathbf{x}_i = \sum_{k=1}^K \lambda_i^k \mathbf{x}_i^k.
\end{equation}
Naturally, the robust representations are calculated as the weighted combinations of the view-specific representations with the coefficients as the voting weights of views, which is quite intuitive.

By integrating both objectives, the final objective of the collaboration framework can be summarized below:
\begin{equation}
\label{eqn::obj_MVEmbed}
O_{collab}=\sum_{k=1}^K O_{k}+\eta R,
\end{equation}
where $\eta$ is a parameter used to control the weight of the regularization term.

\subsection{Learning the Weights of Views through Attention}
\label{subsec::weight}
The above framework proposes a flexible way to let different views collaborate with each other. In this part, we introduce an attention based approach for learning the weights of views during voting. Our proposed approach is very general, which can automatically learn the weights of views for different nodes by providing a few labeled data for specific tasks. For example, for the node classification task, only a few labeled nodes are required; for the link prediction task, a limited number of links are provided.

Following the recent attention based models for neural machine translation~\cite{bahdanau2014neural}, we define the weight of view $k$ for node $v_i$ using a softmax unit as follows:
\begin{equation}
	\label{eqn::weight_softmax}
	\lambda_i^k=\frac{\exp(\mathbf{z}_k \cdot \mathbf{x}_i^C)} { \sum_{k'=1}^{K} \exp(\mathbf{z}_{k'} \cdot \mathbf{x}_i^C)},
\end{equation}
where $\mathbf{x}_i^C$ is the concatenation of all the view-specific representations of node $v_i$, and $\mathbf{z}_k$ is a feature vector of view $k$, describing what kinds of nodes will consider view $k$ as informative. If $\mathbf{x}_i^C$ and $\mathbf{z}_k$ have a large dot product, meaning node $v_i$ believes that view $k$ is an informative view, then the weight of view $k$ for node $v_i$ will be large based on the definition. Besides, we see that the weights of views for each node are determined by the concatenation of its view-specific representations. Therefore, nodes with large proximities are likely to have similar view-specific representations, and thus focus on similar views. 
Such property is quite reasonable, which allows us to better infer the attentions of different nodes by leveraging their proximities preserved in the learned view-specific representations.

With the above weights as coefficients, different view-specific node representations can be weighted combined to obtain the robust representations according to Eqn.~\ref{eqn::MVEmbed_mstep}. Then we may apply the robust node representations to different predictive tasks, and the voting weights could be automatically learned with the backpropagation algorithm~\cite{rumelhart1988learning} based on the predictive error. Specifically, taking the node classification task as an example, we try to minimize the following objective function with respect to the feature vectors of views $\{\mathbf{z}_k\}_{k=1}^K$:

\begin{equation}
\label{eqn::loss_general_2}
O_{attn}=\sum_{v_i\in{S}}L(\mathbf{x}_i, y_i),
\end{equation}
where $S$ is the set of labeled nodes, $\mathbf{x}_i$ is the robust representation of node $v_i$, $y_i$ is the label of node $v_i$, and $L$ is a specific loss function.
Then the gradient of the objective function with respect to $\{\mathbf{z}_k\}_{k=1}^K$ can be calculated as follows:
\begin{equation}
	\label{eqn::derivatives}
	\frac{\partial O_{attn}}{\partial\mathbf{z}_k}=\sum_{v_i\in{S}} \left [ \sum_{l=1}^K (\frac{\partial O_{attn}}{\partial \lambda_i^k} - \frac{\partial O_{attn}}{\partial \lambda_i^l})\lambda_i^k\lambda_i^l \right ] \mathbf{x}_i.
\end{equation}

After optimizing the parameter vectors $\{\mathbf{z}_k\}_{k=1}^K$, the weights of views for both the labeled nodes and unlabeled nodes can be directly calculated with the definition Eqn.~\eqref{eqn::weight_softmax}. In the experiments, we will show that our weight learning method only requires a small number of labeled data to converge (Sec.~\ref{subsec::labeled_data}), and we will also show that our learning method is very efficient (Sec.~\ref{subsec::efficiency}).

In this paper, we investigate two predictive tasks: node classification and link prediction. For the node classification task, the objective function Eqn.~\eqref{eqn::loss_general_2} is defined as the square loss:
\begin{equation}
	\label{eqn::loss_class}
	O_{attn}^{class}=\sum_{v_i \in S} || \mathbf{w}\mathbf{x}_i-\mathbf{y}_i ||_2^2.
\end{equation}
In the objective function, $\mathbf{y}_i$ is the label vector of node $v_i$, in which the dimension $j$ is set as 1 if $v_i$ belongs to category $j$ and set as 0 otherwise. $\mathbf{w}$ is the parameter set of the classifier. 
For the link prediction task, the labeled data are a collection of links and the pairwise loss is used:
\begin{equation}
	\label{eqn::loss_link}
	O_{attn}^{link}=-\sum_{(v_i,v_j) \in S} \cos(\mathbf{x}_i, \mathbf{x}_j).
\end{equation}
In the objective function, $(v_i, v_j)$ is a linked node pair and $\cos(\cdot,\cdot)$ is the cosine similarity between vectors.

\subsection{Model Optimization}

The objective function of our approach can be efficiently optimized with the coordinate gradient descent algorithm~\cite{wright2015coordinate} and the backpropagation algorithm~\cite{rumelhart1988learning}.
Specifically, in each iteration, we first follow existing studies~\cite{tang2015line,tang2015pte} to sample a set of edges from the network, and optimize the view-specific node representations. Then we infer the parameter vectors of views with the labeled data, and update the voting weights of views for different nodes. Finally, different view-specific node representations will be integrated to vote for the robust representations based on the learned weights.
The overall optimization algorithm is summarized in Alg.~\ref{algo::MVEmbed}.

\begin{algorithm}
    \caption{Optimization Algorithm of MVE.}
    \label{algo::MVEmbed}
    \begin{algorithmic}[1]
        \Require \small{$G=(V,E_1,E_2,\ldots,E_K)$, a set of labeled data $S$, number of samples $T$, number of negative samples $N$.}
        \Ensure \small{Robust node representation.}
        \While{not converge}
            \State $\boxdot$ \textbf{\emph{Updating the view-specific node representations.}}
            \While{ smp $\leq$ $T$}		
				\State Randomly pick up a view, denoted as $k$.
				\State Sample an edge from $E_k$ and also $N$ negative edges.
				\State Update view-specific representations w.r.t. Eqn.~\eqref{eqn::obj_MVEmbed}~\eqref{eqn::loss_general_2}.
				\State Update the context representations w.r.t. Eqn.~\eqref{eqn::obj_MVEmbed}.
		    \EndWhile
		    \State $\boxdot$ \textbf{\emph{Updating the voting weights of views for different nodes.}}
            \State Optimize the parameters of the softmax unit w.r.t. Eqn.~\eqref{eqn::loss_general_2}.
		    \State Update the weights of views for each node according to Eqn.~\eqref{eqn::weight_softmax}.
		    \State $\boxdot$ \textbf{\emph{Updating the robust node representations.}}
		    \State Vote for the robust representations according to Eqn.~\eqref{eqn::MVEmbed_mstep}.
        \EndWhile       
    \end{algorithmic}
\end{algorithm}

\subsection{Time Complexity}
The time complexity of the proposed algorithm is determined by three processes: learning the view-specific representations, learning the robust representations and learning the voting weights of views. According to the previous study~\cite{tang2015line}, learning view-specific representations takes $O(|E|dN)$ time, where $|E|$ is the total number of edges in different views, $d$ is the dimension of the node representations, and $N$ is the number of samples in negative sampling. 
Learning robust representations takes $O(|V|dK)$ time, where $K$ is the number of views in a network. Updating voting weights takes $O(|S|dK)$ time, where $|S|$ is the number of labeled data. 
In practice, we only have a very small number of labeled data, and thus $|S| \ll |V|$. Besides, we also have $|V| \ll |E|$ for most networks. Therefore, the total time complexity of our algorithm can be simplified as $O(|E|dN)$, which is proportional to the total number of edges in the given network. 
For most real-world networks, as the number of edges is usually small, our approach will be very efficient in most cases. We will study the efficiency performance of the proposed approach in Sec.~\ref{subsec::efficiency}.

\section{Experiment}

\begin{table*}[!htb]
	\caption{Statistics of the datasets}
	\label{tab::dataset}
	\centering
	\scalebox{0.7}{
		\begin{tabular}{  |c|c|c|c c c c c c|c|} \hline
			\multirow{2}{*}{\textbf{\large{Category}}} & \multirow{2}{*}{\textbf{\large{Dataset}}} & \multirow{2}{*}{\textbf{\large{\# Node}}} & \multicolumn{6}{c|}{\multirow{2}{*}{\textbf{\large{\# Edges in Each View}}}} & \multirow{2}{*}{\textbf{\large{\# Labeled Data}}} \\
			& & & & & & & & & \\ \hline
			\multirow{6}{*}{\large{Node Classification}} & \multirow{2}{*}{\large{DBLP}} & \multirow{2}{*}{\large{69,110}} & 430,117 & 763,029  & 691,090 & & & & \multirow{2}{*}{\large{200}} \\
			& & &(Co-author)&(Citation)&(Text-sim)& & & &  \\ \cline{2-10}
			
			&\multirow{2}{*}{\large{Flickr}} & \multirow{2}{*}{\large{35,314}} & 3,017,530 & 3,531,300  &  & & & & \multirow{2}{*}{\large{100}} \\
			&  & &(Friendship)&(Tag-sim)& & & & &  \\ \cline{2-10}
			
			&\multirow{2}{*}{\large{PPI}} & \multirow{2}{*}{\large{16,545}} &659,781 &14,167 &137,930 &246,274 &1,339 &39,220 & \multirow{2}{*}{\large{200}} \\
			& & &(Coexpression) &(Cooccurrance) &(Database) &(Experimental) &(Fusion) &(Neighbor) & \\ \hline
			
			\hline
			
			\multirow{4}{*}{\large{Link Prediction}} & \multirow{2}{*}{\large{Youtube}}&\multirow{2}{*}{\large{14,901}} &1,940,806 &5,574,249 &2,239,440 &3,797,635 & & &\multirow{2}{*}{\large{500}} \\
			& & &(Friends) &(Subscriptions) &(Subscribers) &(Videos) & & & \\ \cline{2-10}
			
			&\multirow{2}{*}{\large{Twitter}} & \multirow{2}{*}{\large{304,692}} & 35,254,194 & 2,845,120  & 93,051,769 & & & & \multirow{2}{*}{\large{500}} \\
			& & &(Mention)&(Reply)&(Retweet)& & & &  \\ \hline
			
		\end{tabular}
	}
\end{table*}

We evaluate our proposed approach on two tasks including node classification and link prediction. We first introduce our setup.

\subsection{Experiment Setup}

\subsubsection{Datasets}
We select the following five networks, in which the first three are used for the node classification task and the last two for the link prediction task.
\begin{itemize}[leftmargin=*,noitemsep,nolistsep]
	
\item \textbf{DBLP}: An author network from the DBLP dataset~\cite{tang2008arnetminer}\footnote{~\url{https://aminer.org/AMinerNetwork}}. Three views are identified including the co-authorship, author-citation and text-similarity views. The weights of the edges in the co-authorship view are defined as the number of papers coauthored by each pair of authors; the weights in the author-citation view are defined as the number of papers written by one author and cited by the other; the text-similarity view is a 5-nearest neighbor graph and the similarity is calculated based on the titles and abstracts of each author using TF-IDF. For node classification, we select eight diverse research fields as labels including ``machine learning'', ``computational linguistics'', ``programming language'', ``data mining'', ``database'', ``system technology'', ``hardware'' and ``theory''. For each field, several representative conferences are selected, and only papers published in these conferences are kept to construct the three views.

\item \textbf{Flickr}: A user network constructed from Flickr dataset~\cite{Wang-etal12}\footnote{~\url{http://dmml.asu.edu/users/xufei/datasets.html}}, including the friendship view and the tag-similarity view. The tag-similarity view is a 100-nearest neighbor graph between users and the user similarity is calculated according to their tags. The community membership are used as classification labels.
\item \textbf{PPI}: A protein-protein interaction network constructed from the STRING database v9.1~\cite{franceschini2013string}. Only the human genes are kept as nodes. Six views are constructed based on the coexpression, cooccurrance, database, experiments, fusion and neighborhood information. As the original network is very sparse, we follow the same way in~\cite{tang2015line} to reconstruct the six views to make them denser. More specifically, for each view, we expand the neighborhood set of the nodes whose degree are less than 1,000 by adding their neighbors of neighbors until the size of the extended neighborhood set reaches 1,000. The gene groups provided in the Hallmark gene set~\cite{liberzon2011molecular} are treated as the categories of nodes.
\item \textbf{Youtube}: A user network constructed from~\cite{Zafarani+Liu:2009}\footnote{~\url{http://socialcomputing.asu.edu/datasets/YouTube2}}. Five views are identified including the number of common friends, the number of common subscriptions, the number of common subscribers, the number of common favorite videos, and the friendship. We believe the friendship view can better reflect the proximity between the users. Therefore, we select the other four views for training and predict the links in the friendship view.
\item \textbf{Twitter}: A user network constructed from Higgs Twitter Dataset~\cite{de2013anatomy}\footnote{~\url{https://snap.stanford.edu/data/higgs-twitter.html}}. Due to the sparsity of the original network, we treat it as an undirected network here. Four views are identified including reply, retweet, mention and friendship. Similarly, the friendship view is used for link prediction and the other three views are used for training, which are reconstructed in the same way as the PPI dataset.
\end{itemize}

The detailed statistics of these networks are summarized in Table~\ref{tab::dataset}.

\begin{table*} [!htb]
	\caption{Quantitative results on the node classification task. Without learning the weights of views (MVE-NoAttn), our approach has already outperformed all baseline approaches. By learning the weights of views through the attention based approach (MVE), the results are further improved. Removing the collaboration of views (MVE-NoCollab) decreases the results.
	}
	\label{tab::results-classification}
	\begin{center}
		\scalebox{1}{
		\begin{tabular}{|C{2cm}|C{2.5cm}|c c|c c|c c|}\hline
		\multirow{2}{*}{\textbf{Category}}&\multirow{2}{*}{\textbf{Algorithm}}	& \multicolumn{2}{c|}{\textbf{DBLP}} & \multicolumn{2}{|c|}{\textbf{Flickr}} & \multicolumn{2}{|c|}{\textbf{PPI}}\\ \cline{3-8}
		&	&  \textbf{Macro-F1}& \textbf{Micro-F1} &  \textbf{Macro-F1}& \textbf{Micro-F1}  &  \textbf{Macro-F1}& \textbf{Micro-F1} \\ \hline \hline
	\multirow{2}{*}{Single View}&LINE&70.29 &70.77 &34.49 &54.99 &20.69 &24.70  \\  
	&node2vec&71.52 &72.22 &34.43 &54.82 &21.20 &25.04  \\ \hline
	\multirow{8}{*}{Multi View}
	&node2vec-merge &72.05 &72.62 &29.15 &52.08 &21.00 &24.60 \\
	&node2vec-concat &70.98 &71.34 &32.21 &53.67 &21.12 &25.28 \\
	&CMSC&- &- &- &- & 8.97& 13.10\\ 
	&MultiNMF&51.26 &59.97 &18.16 &51.18 &5.19 &9.84\\
	&MultiSPPMI &54.34 &55.65 &32.56 &53.80 &20.21 &23.34 \\ \cline{2-8}
	&MVE-NoCollab&71.85 &72.40 &28.03 &54.62 &18.23 &22.40 \\ 
	&MVE-NoAttn&73.36 &73.77 &32.41 &54.18 &22.24 &25.41\\ 
	&MVE&\textbf{74.51} &\textbf{74.85} &\textbf{34.74} &\textbf{58.95} &\textbf{23.39} &\textbf{26.96} \\ \hline						
	\end{tabular}
	}
	\end{center}
\end{table*}

\begin{table} [!htb]
	\caption{Quantitative results on the link prediction task. MVE achieves the best results through the collaboration framework and the attention mechanism.}
	\label{tab::results-link}
	\begin{center}
		\scalebox{1}{
		\begin{tabular}{|C{2cm}|C{2.5cm}|c|c|}\hline
		\textbf{Category}	&\textbf{Algorithm} & \textbf{Youtube} & \textbf{Twitter}\\ \hline \hline
	 \multirow{2}{*}{Single View}&LINE&85.31&64.18  \\ 
	 &node2vec &88.71 &78.75 \\ \hline
	\multirow{8}{*}{Multi View}
	&node2vec-merge &90.31 &81.80 \\
	&node2vec-concat &92.12 &75.00 \\
	&CMSC&74.25 &-  \\ 
	&MultiNMF&68.30 &-  \\ 
	&MultiSPPMI&86.35 &53.95 \\ \cline{2-4}
	&MVE-NoCollab&89.47 &73.26  \\ 
	&MVE-NoAttn&93.10 &82.62  \\ 
	&MVE&\textbf{94.01} &\textbf{84.98}  \\ \hline				
		\end{tabular}
	}
	\end{center}
\end{table}

\subsubsection{Compared Algorithms}
We compare two types of approaches: single-view based and multi-view based.
\begin{itemize}[leftmargin=*,noitemsep,nolistsep]
    \item \textbf{LINE}~\cite{tang2015line}: A scalable network embedding model for single views. We report the best results on single views.
    \item \textbf{node2vec}~\cite{grovernode2vec}: Another scalable network embedding model for single views. The best results on single views are reported.
    \item \textbf{node2vec-merge}: A variant of the node2vec model. To exploit multiple views of a network, we merge the edges of different views into a unified view and embed the unified view with node2vec.
    \item \textbf{node2vec-concat}: A variant of the node2vec model. To exploit multiple views of a network, we first apply node2vec to learn node representations on each single view, and then concatenate all learned representations.
	\item \textbf{CMSC}: A co-regularized multi-view spectral clustering model~\cite{kumar2011co}, which can apply to our problem but cannot scale up to very large networks. The centroid based variant is used due to its better efficiency, and the centroid eigenvectors are treated as the node representations.
	\item \textbf{MultiNMF}: A multi-view non-negative matrix factorization model~\cite{liu2013multi}, which can also apply to our problem but cannot scale up to very large networks.
	\item \textbf{MultiSPPMI}: SPPMI~\cite{levy2014neural} is a word embedding model, which learns word embeddings by factorizing the word co-occurrence matrices. We leverage the model to learn node representations by jointly factorizing the adjacency matrices of different views and sharing the node representations across different views.
	\item \textbf{MVE}: Our proposed approach for multi-view network embedding, which deploys both the collaboration framework and the attention mechanism.
	\item \textbf{MVE-NoCollab}: A variant of MVE. We introduce different context node representations for different views, so that the view-specific representations will locate in different semantic spaces, and thus they cannot collaborate with each other during training.
	\item \textbf{MVE-NoAttn}: A variant of MVE. We assign equal weights to different views during voting, without learning the voting weights of views through the attention based approach.
\end{itemize}
Note that the DeepWalk model~\cite{perozzi2014deepwalk} can be viewed as a variant of the node2vec model~\cite{grovernode2vec} with the parameters $p$ and $q$ as 1, and thus we will not compare with DeepWalk in our experiments.

\subsubsection{Parameter Settings}
For all approaches except node2vec-concat, the dimension of the node representations is set as 100. For node2vec-concat, the dimension is set as $100K$, and $K$ is the number of views in a network.
For LINE and MVE, the number of negative samples $N$ is set as 5, and the initial learning rate is set as 0.025, as suggested in~\cite{mikolov2013distributed,tang2015line}. 
For node2vec, we set the window size as 10, the walk length as 40, as suggested in~\cite{perozzi2014deepwalk,grovernode2vec}. The parameters $p$ and $q$ are selected based on the labeled data.
For MVE, the number of samples $T$ used in each iteration is set as 10 millions, and the parameter $\eta$ is set as 0.05 by default.

\subsection{Quantitative Results}

\subsubsection{Node Classification}

We start by introducing the results on the node classification task. We treat the node representations learned by different algorithms as features, and train one-vs-rest linear classifiers using the LibLinear package \cite{fan2008liblinear} \footnote{\url{https://www.csie.ntu.edu.tw/~cjlin/liblinear/}}. For both the DBLP and PPI datasets, 10\% nodes are randomly sampled as the training examples and the rest 90\% for testing. For the Flickr dataset, 100 nodes are randomly sampled for training and 10000 nodes for testing. To learn the weights of views, we select a small number of nodes as labeled data, and the concrete numbers are reported in Table~\ref{tab::dataset}.

We present the results of different approaches on the node classification task in Table~\ref{tab::results-classification}. As CMSC cannot scale up to very large networks, only the results on the PPI network are reported. For the single-view based approaches, both LINE and node2vec do not perform well. To leverage the information from multiple views, node2vec-merge combines the edges of different views. However, the proximities of different views are usually not comparable, and simply combining them will destroy the network structures of individual views. On the Flickr dataset, the performance of node2vec-merge is even inferior to the performance of single-view based approaches. On the other hand, node2vec-concat will concatenate all node representations learned on individual views. However, the representations from some sparse views can be very biased, which may destroy the final representations, and thus node2vec-concat does not significantly outperform other approaches, even with much higher dimensions. Both the multi-view clustering method (CMSC) and multi-view matrix factorization methods (MultiNMF and MultiSPPMI) fail to perform well, as they cannot effectively achieve the collaboration of different views and also learn their weights.

For our proposed framework MVE, without leveraging the labeling information to learn the voting weights of views (MVE-NoAttn), it already outperforms all baseline approaches, including node2vec-concat, which learns representations with much higher dimensions. By learning the weights of views using the attention mechanism (MVE), the results are further improved. Besides, if we remove the collaboration of different views (MVE-NoCollab), we observe inferior results, which shows that our collaboration framework can indeed improve the performances by promoting the collaboration of views.

\subsubsection{Link Prediction}

Next we introduce our results on the link prediction task, which aims to predict the links that are most likely to form given existing networks. As the node sets are very large, predicting links on the whole node sets is unrealistic. Therefore, we follow the experimental setting in~\cite{liben2007link} to construct a core set of nodes for each dataset, and we only predict the links between the nodes in the core sets. For the Youtube dataset, the core set contains all the nodes appearing in the four views, which has 7,654 nodes in total. For the Twitter dataset, as there are too many nodes, we randomly sample 5,000 nodes appearing in all the three views. For each pair of nodes in the core set, the probability of forming a link between them is measured as the cosine similarity between their robust node representations. To learn the voting weights of views in our MVE model, we randomly sample 500 edges from each dataset as the labeled data, which are then excluded during evaluation. The performance is measured with the commonly used AUC metric~\cite{fawcett2006introduction}. The results of link prediction with different models are presented in Table~\ref{tab::results-link}. For CMSC and MultiNMF, as they cannot scale up to very large networks, only the results on the Youtube dataset are reported.

We see that for the single view based approaches, both node2vec and LINE fail to perform well. By merging the edges of different views, the results of node2vec-merge are significantly improved, as different views are comparable and complementary on these two datasets. Concatenating the representations learned on each view (node2vec-concat) leads to inferior results on the Twitter dataset, as some sparse views, e.g., the view constructed with the replying relationship, may destroy the concatenated representations. The multi-view clustering method (CMSC) and multi-view matrix factorization methods (MultiNMF and MultiSPPMI) still fail to perform well as they cannot effectively achieve the collaboration of different views. 

For our proposed framework MVE, it outperforms all the baseline approaches. If we remove the collaboration of views (MVE-NoCollab) or remove the weight learning method (MVE-NoAttn), the results will drop, which demonstrates the effectiveness of our collaboration framework and the importance of the attention mechanism.

\subsection{Performances w.r.t. Data Sparsity}

Based on the above results, we have already seen that our proposed approach MVE can effectively leverage the information from multiple views to improve the overall performances. In this part, we take a further step and examine whether MVE is robust to data sparsity by integrating information from multiple views.

\begin{figure}[htb!]
	\centering
	\subfigure[DBLP]{
		\label{fig::dgr_dblp}
		\includegraphics[width=0.22\textwidth]{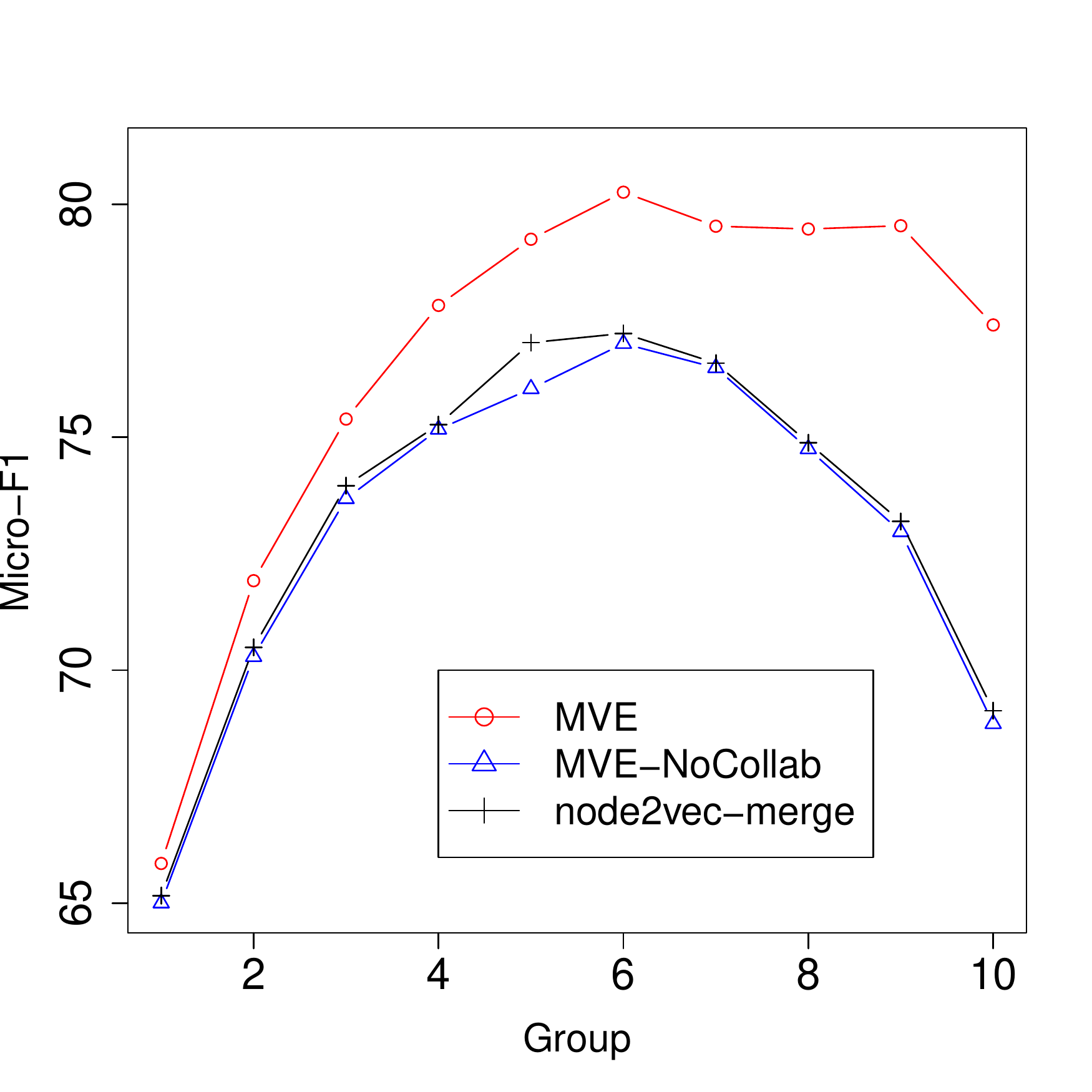}	
	}
	\subfigure[Youtube]{
		\label{fig::dgr_youtube}
		\includegraphics[width=0.22\textwidth]{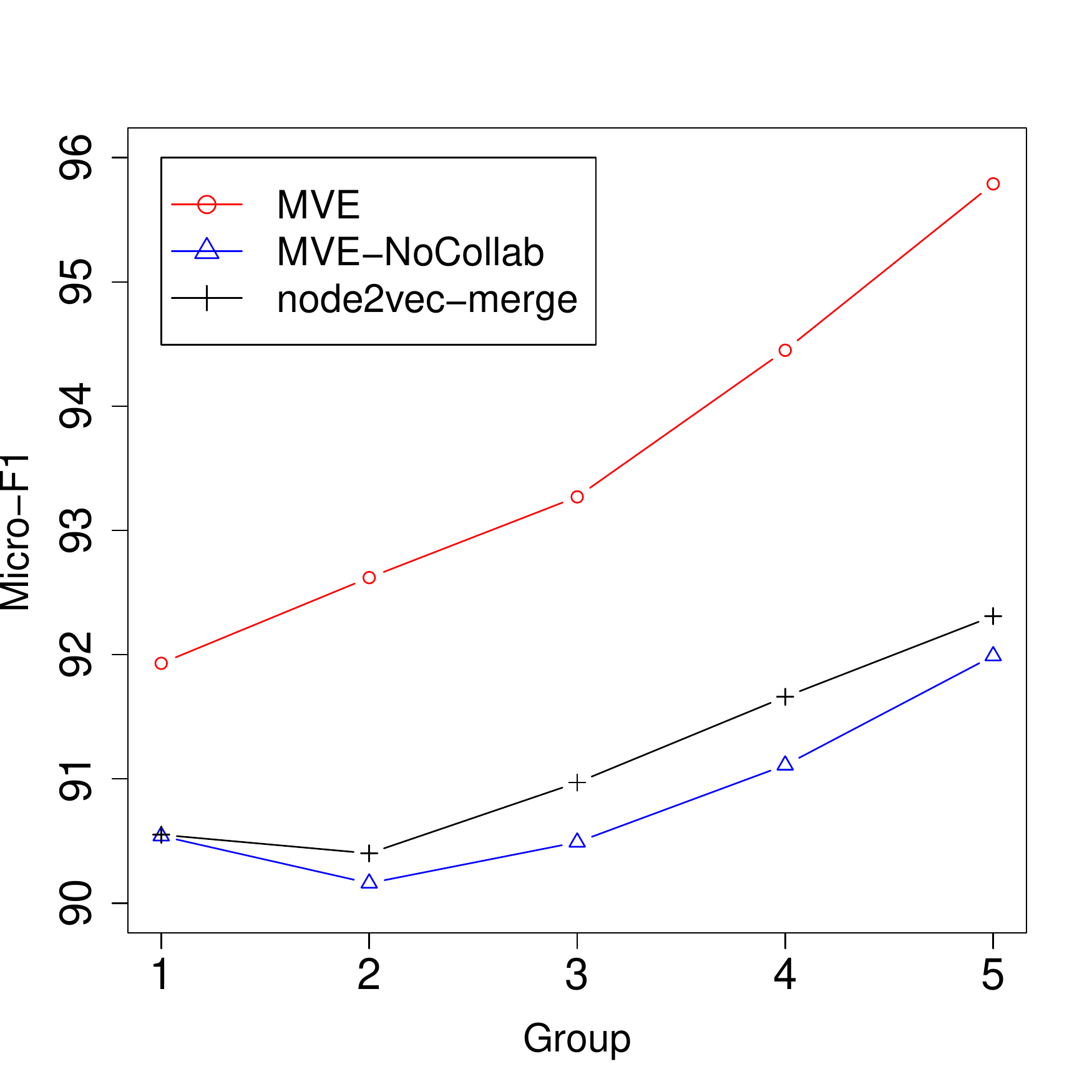}	
	}
	\caption{Performances of the robust node representations w.r.t. sparsity. The left groups consist of high-degree nodes (dense) while the right ones consist of low-degree nodes (sparse). MVE outperforms MVE-NoCollab and node2vec-merge, especially on the right groups (more sparsity).}
	\label{fig::robust-comparison}
\end{figure}

Specifically, we study the performances of MVE on nodes with different degrees, which correspond to different levels of data sparsity. The degree of each node is calculated as the sum of the degrees in different views. All the nodes are assigned into 10 different groups in the DBLP dataset and 5 different groups in the Youtube dataset according to their degrees. We compare the robust node representations learned by MVE, node2vec-merge and MVE-NoCollab (the variant of MVE without promoting the collaboration of views), and we report the performances on different node groups. The results are presented in Figure~\ref{fig::robust-comparison}. The left groups contain nodes with larger degrees, in which the data are quite dense; while the right groups contain nodes with smaller degrees, and the data are very sparse. For the DBLP dataset, all the three models do not perform well on the left node groups since many high-degree nodes belong to multiple research domains, which are more difficult to classify. On the right groups, node2vec-merge and MVE-NoCollab still have quite poor performances, while our proposed MVE significantly outperforms them. For the Youtube dataset, similar results are observed. On the left groups, the performance of the three models are pretty close. On the right groups with low-degree nodes, MVE outperforms both node2vec-merge and MVE-NoCollab.
Overall, compared with MVE-NoCollab and node2vec-merge, we see that MVE achieves better results especially on the right node groups (more sparsity), which demonstrates that our approach can effectively address the data sparsity problem and help learn more robust node representations.

\subsection{Analysis of the Learned Attentions (Weights) Over Views}

In our proposed MVE model, we adopt an attention based approach to learn the weights of views during voting, so that different nodes can focus most of their attentions on the most informative views. The quantitative results have shown that MVE achieves better results by learning attentions~\footnote{The term attention and the term weight have the same meaning here.} over views. In this part, we will examine the learned attentions to understand why it can help improve the performances.

\begin{figure}[htb!]
	\centering
	\subfigure[DBLP]{
		\label{fig::avgwt_dblp}
		\includegraphics[width=0.22\textwidth]{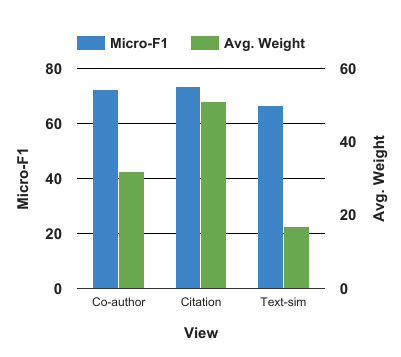}	
	}
	\subfigure[Youtube]{
		\label{fig::avgwt_youtube}
		\includegraphics[width=0.22\textwidth]{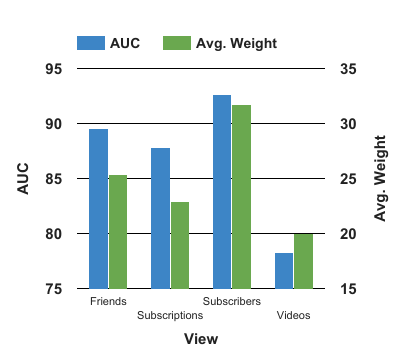}	
	}
	\caption{Comparison of performances on individual views and the average weights of views. Views with better performances usually attract more attentions from nodes.}
	\label{fig::average-weight}
\end{figure}

We first study what kinds of views turn to attract more attentions from nodes. We take the DBLP and Youtube datasets as examples. For each view, we report the results of the view-specific representations corresponded to this view, and also the average attentions assigned by different nodes on this view. The results are presented in Figure~\ref{fig::average-weight}. Overall, the performances of views and the average attentions they receive are positively correlated. In other words, our approach will let different nodes focus on the views with the best performances, which is quite reasonable.

\begin{figure}[htb!]
	\centering
	\includegraphics[width=0.5\textwidth]{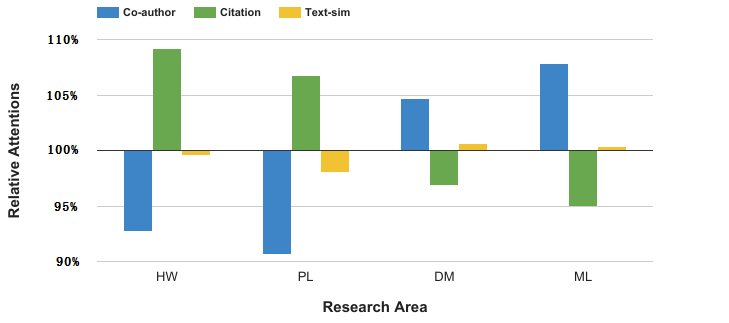}
	\caption{Case study of the learned attentions on the DBLP dataset. We compare the attentions of authors in four research areas. HW stands for hardware, PL for programming language, DM for data mining and ML for machine learning.}
	\label{fig::case-weights}
\end{figure}

We further study the learned attentions on a more fine-grained level by comparing the attentions of nodes in different semantic groups. We take the DBLP dataset as an example, and study the authors in four different research areas including hardware (HW), programming language (PL), data mining (DM) and machine learning (ML). For each research area, we calculate the average view weights assigned by authors within this area, and we report the ratio to the average view weights assigned by other authors, in order to know which views are the most informative for each research area. The results are presented in Figure~\ref{fig::case-weights}. For authors in the areas of hardware and programming language, they have relatively more attentions on the author citation view; while for authors in the areas of data mining and machine learning, they focus more on the co-authorship view. This is because we are studying the node classification task, aiming at predicting the research areas of different authors. To increase the prediction accuracy, our attention mechanism needs to let the authors focus on the views that can best discriminate them from authors in other areas.
For authors in the hardware and programming language areas, they may cite very different papers compared with authors in other areas, and hence the author citation view is the most discriminative for them, so they have more attentions on the author citation view. On the other hand, several areas in our dataset are related to artificial intelligence, such as data mining and machine learning. For authors in those areas, they may use similar terms as well as cite similar papers, so the text-similarity view and the author citation view cannot discriminate these areas from each other. Therefore, authors in these areas pay less attentions to the text-similarity view and the author citation view, and they focus more on the co-authorship view.

Overall, the attentions (weights) over views learned by our attention mechanism are very intuitive, which enable different nodes to focus on those most informative views. 

\subsection{Parameter Sensitivity}

Next, we investigate the sensitivity of different parameters in our framework, including $\eta$ and the number of labeled data.

\begin{figure}[htb!]
	\centering
	\subfigure[DBLP]{
		\label{fig::ps_eta_dblp}
		\includegraphics[width=0.22\textwidth]{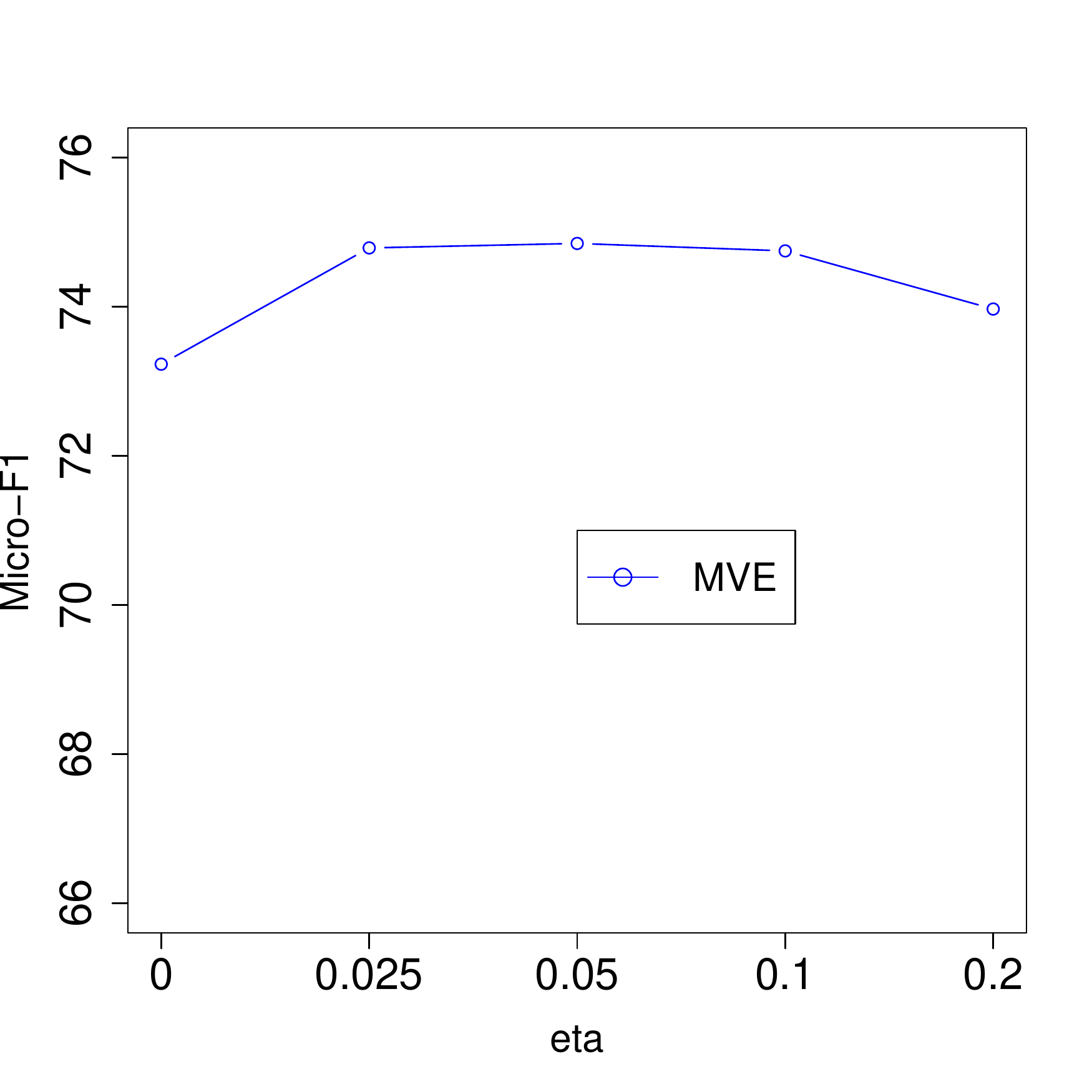}
	}
	\subfigure[Youtube]{
		\label{fig::ps_eta_youtube}
		\includegraphics[width=0.22\textwidth]{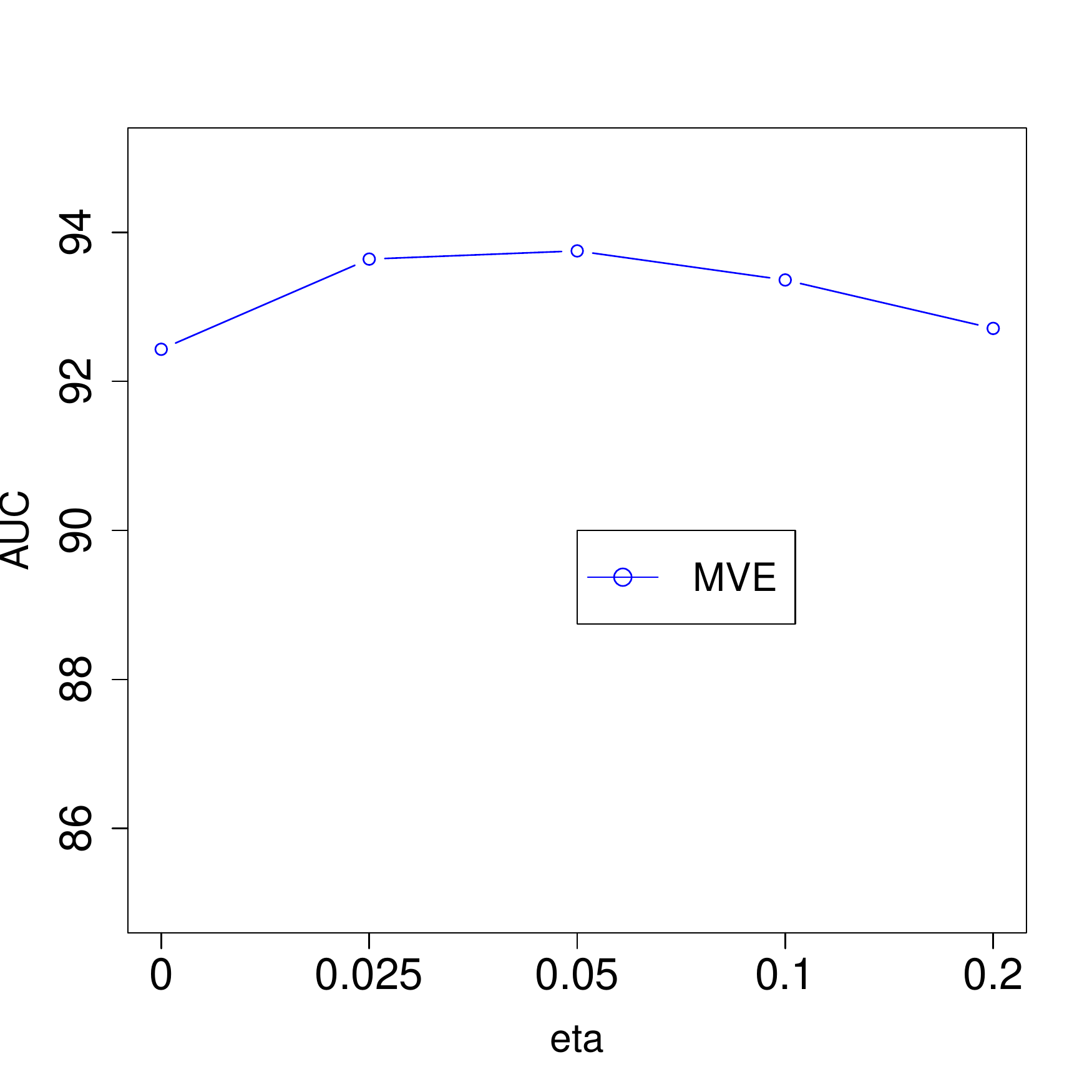}	
	}
	\caption{Performances w.r.t. $\eta$. Within a large range $(0.025,0.1)$, the performance is not sensitive to $\eta$. The performance remains very stable with the default value 0.05.}
	\label{fig::para_sen_eta}
\end{figure}

\begin{figure}[htb!]
	\centering
	\subfigure[DBLP]{
		\label{fig::ps_lb_flickr}
		\includegraphics[width=0.22\textwidth]{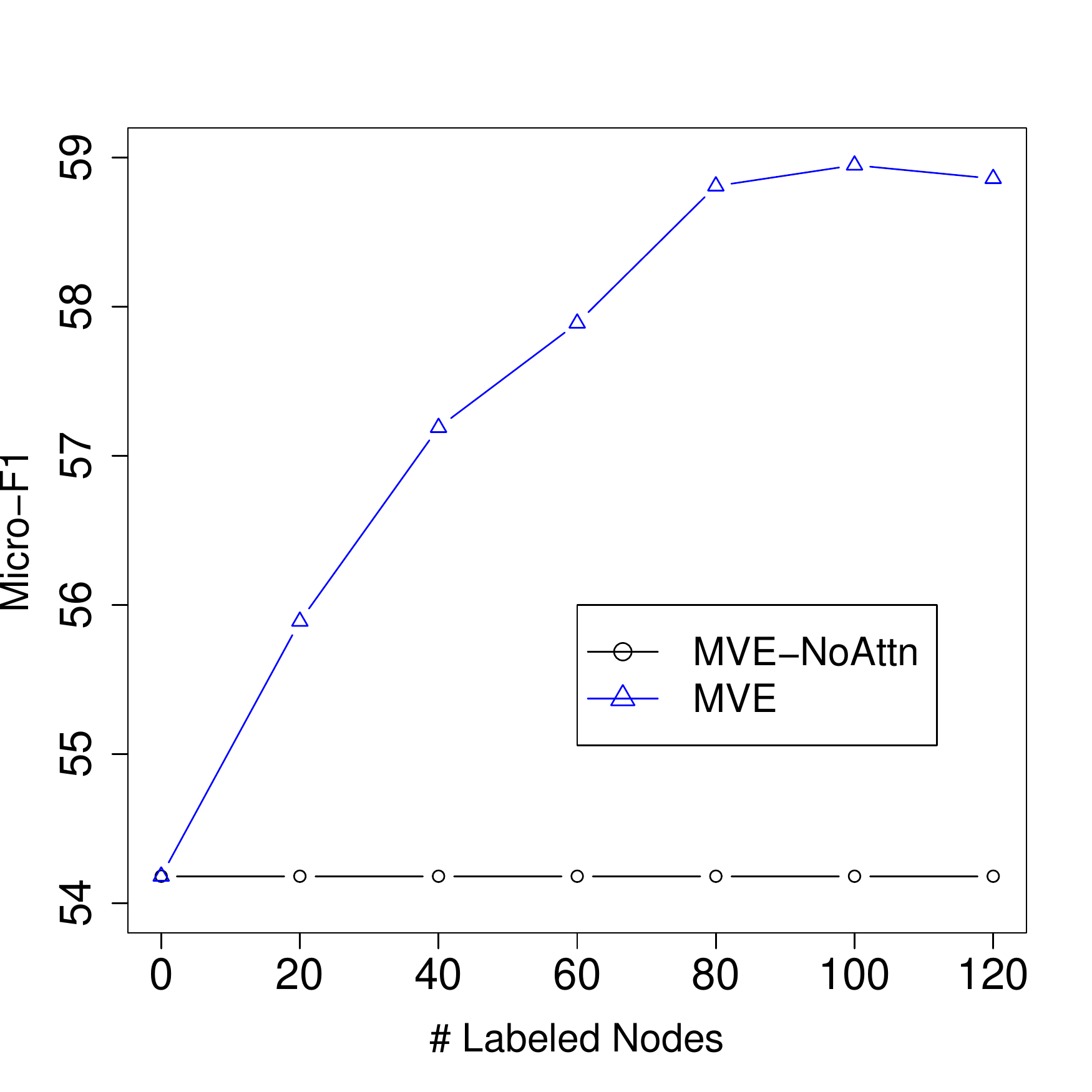}
	}
	\subfigure[Youtube]{
		\label{fig::ps_lb_tweet}
		\includegraphics[width=0.22\textwidth]{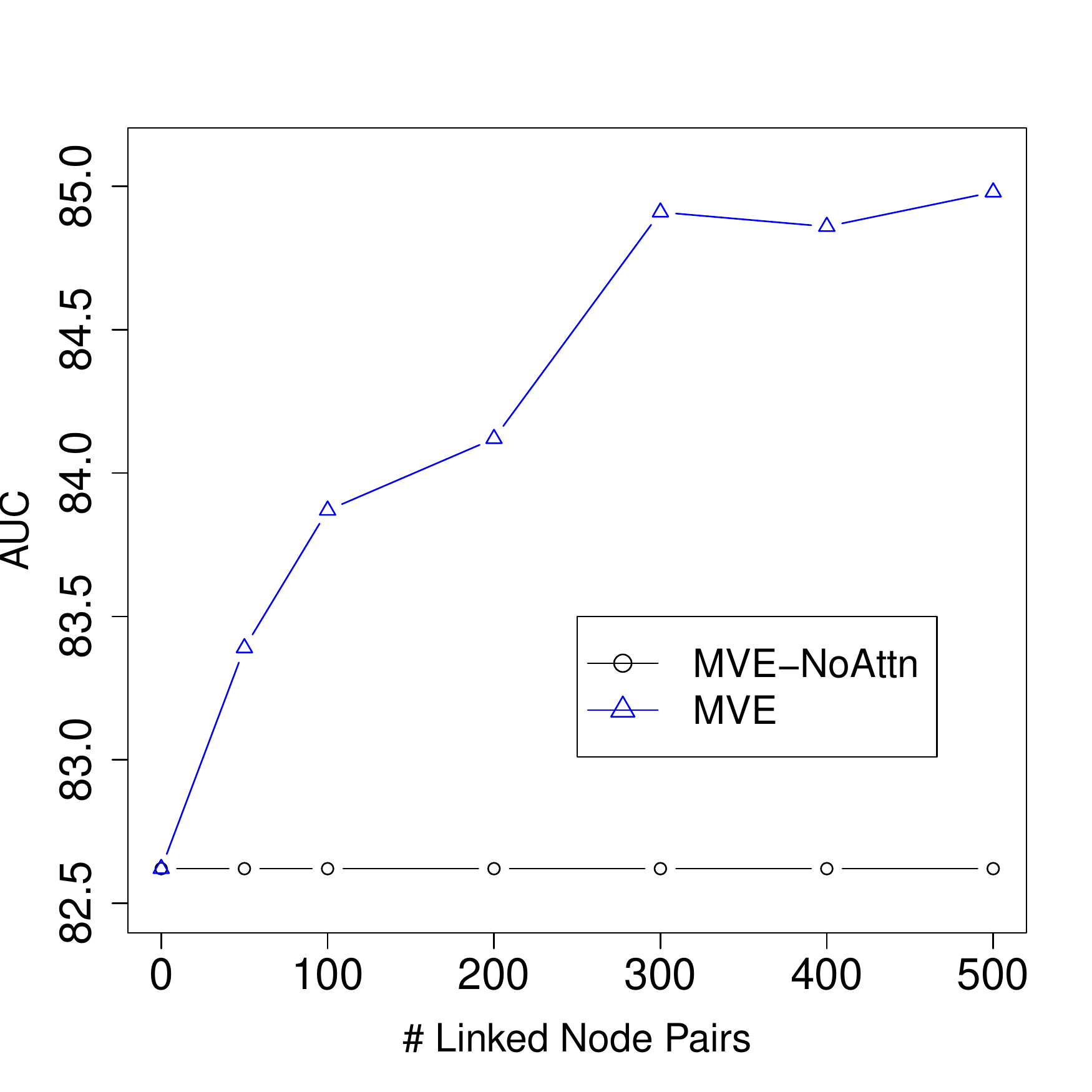}
	}
	\caption{Performances w.r.t. \#labeled data. By learning the voting weights of views with the labeled data, MVE consistently outperforms MVE-NoAttn, which assigns equal weights. MVE requires only a few labeled data to converge.}
	\label{fig::para_sen_lb}
\end{figure}

\subsubsection{Performances w.r.t. $\eta$}
In our collaboration framework, the parameter $\eta$ controls the weight of the regularization term (Eqn.~\ref{eqn::obj_MVEmbed}), which trades off between preserving the proximities encoded in single views and reaching agreements among different views. We compare the performances of our proposed approach w.r.t. $\eta$ on both the node classification task and the link prediction task.

Figure~\ref{fig::para_sen_eta} presents the results on the DBLP and Youtube datasets. When $\eta$ is set as 0, all the three approaches will not perform so well as different views are not able to communicate with each other through the regularization term. As we increase $\eta$ from 0, the performances are improved, which remain stable with a large range $(0.025,0.1)$ on both datasets. If we further increase $\eta$, the performances will begin to drop. This is because a large $\eta$ forces different views to fully agree with each other, ignoring the differences between the views.

\subsubsection{Performances w.r.t. the number of labeled nodes}
\label{subsec::labeled_data}
To learn the voting weights of views for different nodes, our framework requires some labeled nodes. In this part, we investigate the performance of our framework w.r.t. the number of labeled nodes. We take the Flickr and Twitter datasets as examples, and report the performances of both MVE and MVE-NoAttn. 

We present the results in Figure~\ref{fig::para_sen_lb}. We see that by leveraging the labeled nodes to learn the voting weights of views, MVE consistently outperforms its variant MVE-NoAttn, which assigns equal weights to different views. On both datasets, MVE requires only a very small number of labeled nodes to converge, which shows the effectiveness of our attention based method for weight learning.

\subsection{Efficiency Study}
\label{subsec::efficiency}

In this part, we study the efficiency of our proposed framework. We select the DBLP and Twitter datasets as examples, and compare the running time of MVE with node2vec, LINE and MVE-NoAtten (the variant of MVE without learning the voting weights of views).

Table~\ref{table::efficiency} presents the results. We see that MVE has close running time with LINE and node2vec on both datasets. On the Twitter dataset with more than 300 thousands nodes and 100 millions edges, the training process of MVE takes less than 15 minutes, which is quite efficient. Besides, comparing the running time of MVE and MVE-NoAttn, we observe that the weight learning process in MVE takes less than 15\% of the total running time on both datasets, which shows the good efficiency of our attention based approach for weight learning.

\subsection{Case Study}

Our collaboration framework can effectively preserve the node proximities encoded in different views through the view-specific representations, which are further used to vote for the robust node representations. In this part, we give some illustrative examples to show the differences between the view-specific and the robust node representations. We take the author network in DBLP as an example. To compare these node representations, we list the most similar authors given a query author according to the cosine similarity calculated with different node representations. Table~\ref{tab::cases} presents the results. From the nearest neighbors, we can see that the view-specific node representations can well preserve the proximities encoded in the individual views, whereas the robust representations combine the information from all different views.

\begin{table}[t]
    \caption{Efficiency study. Our approach has close running time to LINE and node2vec. Learning weights of views takes less than 15\% of the running time on both datasets.}
    \label{table::efficiency}
    \begin{center}
    \begin{tabular}{  |C{2cm}|C{2cm}|C{2cm}|}
        \hline
        \textbf{Algorithm} & \textbf{DBLP} & \textbf{Twitter} \\ \hline
        LINE & 91.45 s & 589.29 s  \\
        node2vec & 144.77 s & 981.96 s  \\
        \hline
        MVE-NoAttn& 105.05 s & 732.26 s \\
        MVE& 120.38 s & 847.65 s \\
        \hline
        \end{tabular}
    \end{center}
    \vspace{-0.3cm}
\end{table}

\begin{table*}[!htb]
    \vspace{-0.5cm}
	\caption{ Examples of nearest neighbors according to similarity calculated by view-specific node representations and robust node representations on the DBLP dataset. }
	\label{tab::cases}
	\centering
	\scalebox{0.9}{
		\begin{tabular}{ |c | C{3.5cm} | C{3.5cm} | C{3.5cm} | C{3.5cm} |}\hline
			\textbf{Query} & \textbf{View1: Co-authorship} &\textbf{View2: Author-citation} & \textbf{View3: Text-similarity} & \textbf{Robust}\\ \hline
			\multirow{5}{*}{Jiawei Han} &      Hong Cheng & Jian Pei & Philip S. Yu & Jian Pei
			\\ 
			&Xifeng Yan & Rakesh Agrawal & Qiang Yang & Xifeng Yan
			\\ 
			&Jing Gao & Ramakrishnan Srikant & Christos Faloutsos & Philip S. Yu
			\\ 
			&Xiaolei Li & Philip S. Yu & Zheng Chen & Dong Xin
			\\ 
			&Feida Zhu & Ke Wang & S. Muthukrishnan & Hong Cheng
			\\ \hline
			\multirow{5}{*}{Michael I. Jordan} &      Erik B. Sudderth & David M. Blei & David Barber & Erik B. Sudderth
			\\ 
			&Francis R. Bach & Andrew Y. Ng & Ryan Prescott Adams & Christopher M. Bishop
			\\ 
			&David M. Blei & Andrew McCallum & Emily B. Fox & David M. Blei
			\\ 
			&Ling Huang & Thomas Hofmann & David B. Dunson & Yee Whye Teh
			\\ 
			&Tommi Jaakkola & Eric P. Xing & Frank Wood & Alan S. Willsky
			\\ \hline
		\end{tabular}
	}
\end{table*}

\section{Related Work}
\label{sec::related}

Our work is related to the existing scalable approaches for learning network representations including DeepWalk~\cite{perozzi2014deepwalk}, LINE~\cite{tang2015line} and node2vec~\cite{grovernode2vec}, which use different search strategies to exploit the network structures: depth-first search, breadth-first search, and a combination of the two strategies. However, all these approaches focus on learning node representations for networks with a single view while we study networks with multiple views. 

The other line of the related work is multi-view learning, which aims to exploit information from multiple views and has shown effectiveness in various tasks such as classification~\cite{blum1998combining,kumar2011co,wang2010new}, clustering~\cite{kumar2011co,xia2010multiview,zhou2007spectral,chaudhuri2009multi,kumar2011co}, ranking~\cite{yu2014exploiting}, topic modeling~\cite{tang2013one} and activity recovery~\cite{zhang2017regions}. 
The work which is the most similar to ours is the multi-view clustering~\cite{kumar2011co,xia2010multiview,zhou2007spectral,chaudhuri2009multi} 
and multi-view matrix factorization~\cite{liu2013multi,greene2009matrix,singh2008relational} methods.  
For example, Kumar et al.~\cite{kumar2011co} proposed a spectral clustering framework to regularize the clustering hypotheses across different views. 
Liu et al.~\cite{xia2010multiview} proposed a multi-view nonnegative matrix factorization model, which aims to minimize the distance between the coefficient matrix of each view and the consensus matrix. 
Our multi-view network representation approach shares similar intuition with these pieces of work, aiming to find robust data representations across multiple views. 
However, a major difference is that existing approaches assign equal weights to all views, while our approach adopts an attention based method, which learns different voting weights of views for different nodes.

Besides, our work is also related to the attention based models, which aim to infer the importance of different parts of the training data, and let the learning algorithms focus on the most informative parts. Attention based models have been applied to various tasks, including image classification~\cite{mnih2014recurrent}, machine translation~\cite{bahdanau2014neural} and speech recognition~\cite{chorowski2014end}. To the best of our knowledge, this is the first effort to adopt the attention-based approach in the problem of multi-view network representation learning.
\section{Conclusions}
In this paper, we studied learning node representations for networks with multiple views. We proposed an effective framework to let different views collaborate with each other and vote for the robust node representations across different views. 
During voting, we proposed an attention based approach to automatically learn the voting weights of views, which requires only a small number of labeled data.
We evaluated the performance of our proposed approach on five real-world networks with multiple views. 
Experimental results on both the node classification task and link prediction task demonstrated the effectiveness and efficiency of our proposed framework.
In the future, we plan to apply our framework to more applications. 
One promising direction is learning node representations for heterogeneous information networks, i.e., networks with multiple types of nodes and edges. 
In such networks, each meta-path~\cite{sun2011pathsim} characterizes a type of proximity between the nodes, and various meta-paths yield networks with multiple views.

\section*{Acknowledgments}
Research was sponsored in part by the U.S. Army Research Lab. under Cooperative Agreement No. W911NF-09-2-0053 (NSCTA), National Science Foundation IIS-1320617 and IIS 16-18481, and grant 1U54GM114838 awarded by NIGMS through funds provided by the trans-NIH Big Data to Knowledge (BD2K) initiative (www.bd2k.nih.gov). The views and conclusions contained in this document are those of the author(s) and should not be interpreted as representing the official policies of the U.S. Army Research Laboratory or the U.S. Government. The U.S. Government is authorized to reproduce and distribute reprints for Government purposes notwithstanding any copyright notation hereon. Research was partially supported by the National Natural Science Foundation of China (NSFC Grant Nos. 61472006, 61772039 and 91646202).

\bibliographystyle{abbrv}
\bibliography{sigproc}
\end{document}